# Multi-line CO Imaging of Two Ultraluminous Infrared Galaxies

Grishma Adenkar, Viktor Lipovka, Nihar Prabhala & Srikar Vakkalagadda

(Dated: October 26, 2023)

## ABSTRACT

The primary goal of this project was to use the given data of four emission lines from two ultraluminous infrared galaxies to calculate the molecular gas mass and dynamical mass of each galaxy. These quantities can provide valuable information about a galaxy's age, star formation properties, and molecular make-up. Ultraluminous infrared galaxies are formed from the merger or interaction of gas-rich galaxies, and are classified by having a luminosity of greater than $10^{12}$ L$_\odot$. The presence of molecular gas clouds and dust encourages formation of young, bright stars in these galaxies, but also makes it difficult to measure the total gravitational and molecular masses, since molecular hydrogen is virtually undetectable within gas clouds. Our data instead observed emissions of carbon monoxide (CO) from the two ULIRGs at transitions of J= 1→0 orJ= 2→1. The software Difmap was used to map the interferometer data obtained from the Plateau de Bure Interferometer (PdBI). Our data was cleaned, filtered, and subsequently used to generate UV-plots of the emissions, along with other values needed for calculations. Some of the key results include that iras 15250 is much younger as it has not used up a majority of its molecular gas while iras 17208 is older because it has used it up, the gas mass fraction can be used to estimate the amount of dark matter present in the galaxy, and that the gas content and the central surface brightness of the disk are directly correlated.

## 1. INTRODUCTION

Understanding galactic structures and molecular gas mass in galaxies is critical to understanding both star formation and galaxy formation. Obtaining this information is difficult in cases where galaxy resolution is not high enough to distinguish between galactic structures, since the galaxy's molecular makeup cannot be accurately inferred. Ultraluminous Infrared Galaxies, also referred to as ULIRGS, are galaxies with luminosities above $10^{12}$ L$_\odot$ in infrared wavelengths. They were first discovered in large numbers by the Infrared Astronomical Satellite in 1983, and were presumed to form from mergers of gas-rich disks of spiral galaxies. The abundance of molecular gas clouds in these galaxies is what generates high star formation rates, which is why luminous galaxies have very large populations of young, massive stars. The bright luminosity is generally attributed the active galactic nucleus (AGN) at the center of a ULIRG, along with the multitude of bright, young stars. ULIRGs are rich in gas and dust, which together absorb light from stars in the form of Ultraviolet photons, and re-emit it in the infrared.

Even though molecular hydrogen (H$_2$) is the most abundant molecule in the universe, the presence of molecular gas clouds renders it difficult to measure the amount of it in ULIRGs. Only quadrupole rotational transitions can occur with molecular hydrogen, but these are high energy transitions and therefore only occur at higher temperatures (around 1200 K). The vast majority of molecular hydrogen within a galaxy is at a colder temperature of around 30 K, and since there are no transitions at this temperature there are no emitted photons to detect. Carbon monoxide (CO) on the other hand is the second most abundant molecule in galaxies, and it has a dipolar moment that has an easily observable transition ($J = 1 \rightarrow 0$) that occurs at lower temperatures. Since CO and H$_2$ are formed under similar conditions, astronomers typically assume a standard ratio between the number of H$_2$ molecules and the number of CO molecules present in a galaxy. This ratio is typically noted as $10^4$ H$_2$ molecules for every 1 CO molecule, but this value is highly debated within the astronomical community and should not be assumed true for all cases.

Section 2 of this introduces our data and the galaxies observed . Section 3 describes our methodology. Results for individual galaxies are given in Section 4. We discuss our results in Section 5 and conclude in Section 6.

## 2. THE DATA

The data used in this research consists of four pre-calibrated sets of visibilities from two different Ultraluminous infrared galaxies (ULIRGs): infrared source (iras) 15250 and infrared source 17208. These visibilities were obtained with the Plateau de Bure Interferometer (PdBI), which is a facility located in the French Alps and is operated by the Institut de Radio Astronomie Millimétrique (IRAM). The PdBI is now a part of the Northern Extended Millimeter Array (NOEMA) which is also operated by IRAM. The images themselves are carbon monoxide emission lines of either $J = 1 \to 0$ or $J = 2 \to 1$, one of each for each galaxy.

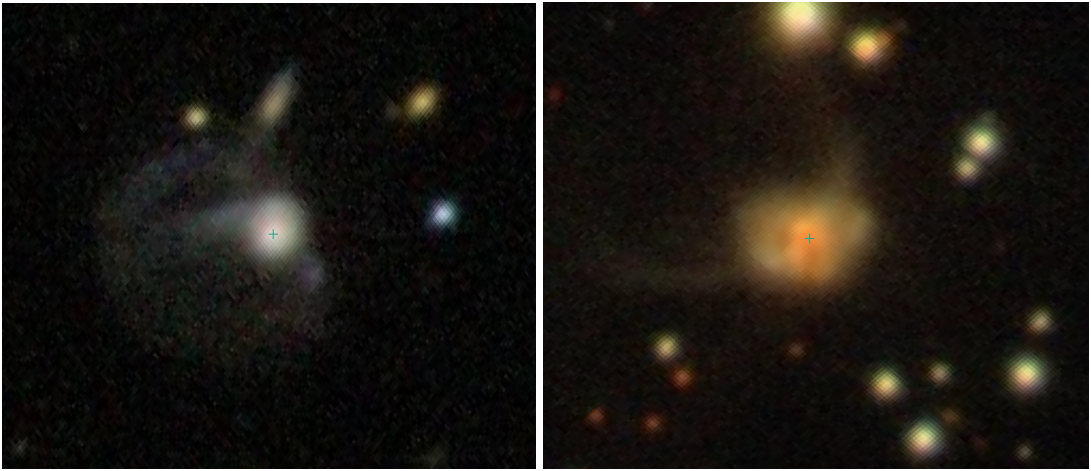

**Figure 1.** Iras 15250 and iras 17208 ULIRGs taken from the Aladin Sky Atlas.

Iras 15250 and iras 17208 both have substantial dust luminosities, which by definition indicates over 100 times that of the Milky Way. These large dust luminosities were created by the bursts of star formation that were triggered by the merger of two gas-rich progenitor galaxies. This is evident in the optical images of the galaxies because of their irregular shapes that are not common spiral or elliptical galaxies.

Figure 2 is an example of the dirty map of the emission as shown in Difmap, the software used in this research.

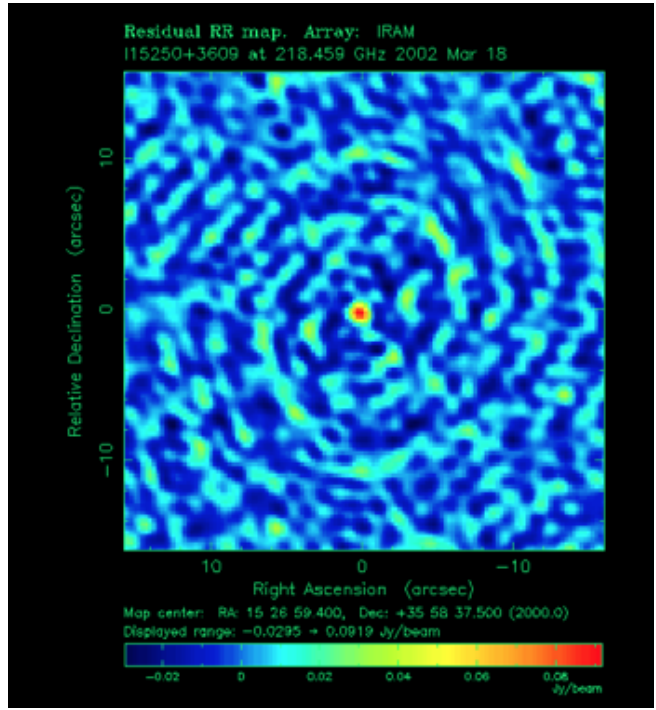

**Figure 2.** Dirty map of iras 17208.

### 3. METHODOLOGY

The core software utilized for the analysis of our data was Difmap. Difmap is an editing and mapping program that allows us to view and map interferometer data in both continuum and spectral lines. After loading our data into Difmap and initializing the mapplot, our first objective was to obtain key properties from the data and identify which frequency bands show emission. The key properties that were obtained included the frequency at the origin (channel 1), the frequency offset between each channel, and the calculated redshift. The frequency at the origin and the frequency offset per channel is given data that Difmap displays, while the redshift can be calculated using the following equation, $1 + z = \frac{\nu_0}{\nu}$, where $\nu_0$ is the given rest frequency and $\nu$ is the frequency at the origin given through Difmap. We were able to identify which frequency bands our emissions was located in by manually sorting through the frequency bins and looking for the emission. Once the central emission was indistinguishable from the surrounding noise, it was determined that there was no longer any emission.

The next objective was to find the range of velocities over which each galaxy shows emission. This was done by using the following equation $\frac{\Delta v_0}{c} = \frac{\Delta \nu}{\nu}$ and solving for the rest-frame velocity interval, $\Delta v_0$. In this equation, $c$ represents the speed of light, $\Delta \nu$ represents the range of frequency that the emission is contained in, and $\nu$ represents the frequency at the peak of the emission. As stated earlier, the peak of the emission is always around the central bin, frequency bin 56. The frequency at this bin was found by utilizing the frequency at the origin (frequency channel 0) and subtracting the offset between each channel 56 times. $\nu_1$ and $\nu_2$ are the edges of the emission, where before and after these points, respectively, there is only noise and no discernible emission line.

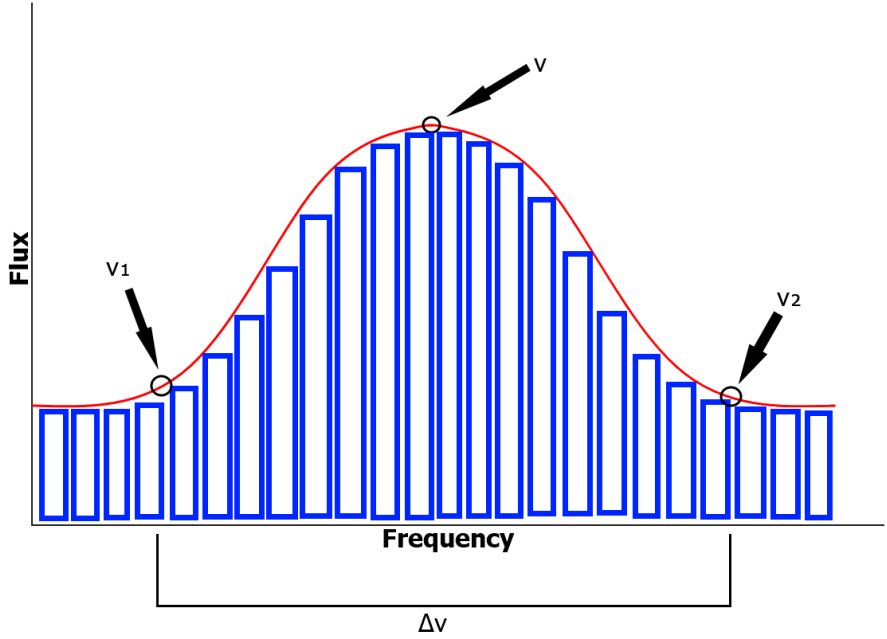

**Figure 3.** Visualization of the frequency bins.

The next objective was to find the spectral line fluxes, $F_{CO(1-0)}$ and $F_{CO(2-1)}$, and then solve for the total molecular gas mass of the galaxy,

$$\frac{M_{H_2}}{M_\odot} = 1.180 \times 10^4 (\frac{D}{Mpc})^2 (\frac{X}{3 \times 10^{20}\ cm^{-2}(K\ km\ s^{-1})^{-1}})(\frac{F_{CO(1-0)}}{Jy\ km\ s^{-1}})$$

where $M_{H_2}$ is the mass of hydrogen in the galaxy, D is the approximate distance to the galaxy, X is a conversion factor between the number of $CO$ and $H_2$ molecules, and $F_{CO(1-0)}$ is the spectral line flux of the J=1-0 line multiplied by the range of velocities over which the emission is present, which was calculated earlier.

$$D \approx \frac{cz}{H_0}$$

$$H_0 \approx 70\ km\ s^{-1}\ Mpc^{-1}$$

$$X \equiv \frac{N_{H_2}}{I_{CO(1-0)}} = 3 \times 10^{20}\ cm^{-2}(K\ km\ s^{-1})^{-1}$$

Using the spectral line fluxes, the flux ratio of the emission lines for both galaxies can be calculated. Next, the total dynamical mass of each galaxy can be solved using the following equation, $M_{dyn} = \frac{Rv^2}{G}$, where $R$ is the radius of the galaxy, $v$ is the velocity of the galaxy, and $G$ is the gravitational constant equal to $6.67 \times 10^{-11} m^3 kg^{-1} s^{-2}$. The radius $R$, was found in several steps. The first step is to manually go through the frequency bins in Difmap and use Difmap to find the location of the peak flux in each set of bins. In most cases, the location of the peak flux will be in the same location as where it is in the peak frequency, but sometimes it is in a slightly different location which indicates a slight difference in angular separation, $\theta$, in arc seconds. Then using our previously calculated redshift, we can use Ned Wright's Cosmology Calculator and find the angular size distance, $D_A$, of the galaxy. Using both $\theta$ and $D_A$, $\Delta R$ is found via the following diagram. Then, $R$ is finally found by multiplying $\Delta R$ by 0.5. Likewise, $v$ is found by multiplying $\Delta v$, the velocity range, by 0.5.

Finally, the total molecular gas mass fraction was solved using the following equation of $f_{gas} \equiv \frac{M_{H_2}}{M_{dyn}}$.

5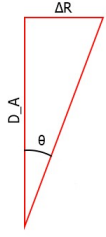

**Figure 4.** Finding $\Delta R$ using $\theta$ and $D_A$.

## 4. RESULTS

Our results can be summarized in the following table.

| $\nu_0 = 115.271\,\text{GHz}$ and $230.538\,\text{GHz}$ | iras 15250 | | iras 17208 | |
|---|---|---|---|---|
| **Emission Line** | J=1→0 | J=2→1 | J=1→0 | J=2→1 |
| **Redshift** | 0.0526 | 0.0524 | 0.0427 | 0.0431 |
| **Velocity Range** | 657.64 km/s | 384.303 km/s | 1461.97 km/s | 719.125 km/s |
| **Spectrum Line Flux** | 0.0125043 Jy | 0.0598549 Jy | 0.0602807 Jy | 0.332033 Jy |
| **Total Molecular Gas Mass** | $4.93*10^9\,M_\odot$ | | $3.48*10^{10}\,M_\odot$ | |
| **Flux Ratio** | 0.2089 (J=1→0 : J=2→1) | | 0.1816 (J=1→0 : J=2→1) | |
| **Total Dynamical Mass** | $6.37*10^9\,M_\odot$ | | $1.05*10^{12}\,M_\odot$ | |
| **Total Molecular Gas Fraction** | 0.774 | | 0.033 | |

**Figure 5.** This table summarizes the results of all the steps outlined in the Methodology section. The most important ones of note are Redshift, Total Molecular Gas Mass, Total Dynamical Mass, and Total Molecular Gas Fraction. It should be noted that the values for $\nu_0$ in the top left box are for $J = 1 \to 0$ and $J = 2 \to 1$, respectively.

As mentioned in the Methodology section, redshift was calculated using the equation $1 + z = \frac{\nu_0}{\nu}$. $\nu$ was the frequency given for the origin of the emission. For both galaxies, we calculated two separate redshifts based on the two separate spectral emission data. For iras 15250, the average of our calculated redshift was 0.0525. The actual observed value is 0.0552 , which gives our calculated redshift an error of about 5%. For iras 17208, the average between the two calculated redshift values is 0.0429; with an actual observed value of 0.042793, our calculated redshift has an error of less than 0.2%. It is encouraging to see that our redshifts are consistent within the same galaxy. Redshift tells us how fast a galaxy is receding away from us. Using the Doppler Redshift calculator, we find that iras 15250 is moving away from us at a velocity of about 5% the speed of light and iras 17208 is moving away from us at a velocity of about 4% the speed of light.

The next important piece of information to note from the table is the differing values for the velocity range. The velocity range was calculated by finding the bounds of the bands where each emission line shows emission. In short, for iras 15250, for example, for the emission line of $J = 1 \to 0$, the frequency bounds showed that the gas that shows this specific emission has a velocity range of about $660 km/s$. To further analyze this, it is important to note exactly what is meant by $J = 1 \to 0$ and $J = 2 \to 1$ exactly means. Emission spectra are the phenomenon that occurs when an atom loses energy. This is specifically defined by the energy of its electrons. An atom in its ground state or $J = 0$, has the lowest possible energy. An atom in an excited state has electron(s) that are in higher energy levels than they would normally be, meaning they have more potential energy than in the ground state. When the excited electron eventually expends that energy and moves back to a lower energy level, by conservation of energy, that energy had to have gone somewhere. That somewhere is electromagnetic radiation, or light that is emitted at specific frequencies, which is what we are observing. What the different velocity ranges tell us about the CO molecules and their behavior is the radial velocity of CO molecules that undergo these specific emissions. It can tell us how fast the galaxy is moving and rotating. It is also important to note that this velocity range is calculated within the frame of the galaxy itself, not our frame as an observer.

Next, we move to the total molecular gas mass calculation. This is arguably the most important result of our analysis. This equation,

$$\frac{M_{H_2}}{M_\odot} = 1.180 \times 10^4 (\frac{D}{Mpc})^2 (\frac{X}{3 \times 10^{20} \ cm^{-2}(K \ km \ s^{-1})^{-1}})(\frac{F_{CO(1-0)}}{Jy \ km \ s^{-1}})$$

, is the crux of our project. The most important term in this equation is the conversion factor, or $X$. This term is what allows us to go from the data we have calculated about the emission of CO molecules, specifically the flux of $CO_{J=1\to 0}$ to a mass for $H_2$ molecules. It is vital to note the history of this conversion factor and assumptions that we have made regarding this term. The conversion factor is an empirically determined term that is generally accepted among the scientific community for calculating the mass of $H_2$ molecules in a system. Of note is that it was calculated for spiral galaxies, so the use of it is valid in our calculations, given that our galaxies are spiral as well. There is some dispute, however, as to whether this conversion factor is totally valid. Both $H_2$ and CO form under very similar, but not the exact same circumstances in space. Given that we do not know exactly the similarities and differences for these circumstances, however, the use of this conversion factor is not 100% accepted. For our purposes, however, we accepted this conversion factor as valid. The crux of our project was to use CO as a tracer molecule for $H_2$, given that $H_2$ is difficult to detect in cold dark clouds of gas among galaxies (it does not particularly exhibit an emission spectrum given that it is a stable molecule without movement of electrons. This is what it means when we say molecular hydrogen does not have a dipole moment; there is no polarization of the molecule due to the movement of the electrons and thus, no emission). Molecular hydrogen is by far the most abundant gas in the universe, so we can learn a lot of about a galaxy by determining the amount of $H_2$ it has.

The next step was determining the total dynamical mass of the galaxy. The dynamical mass is the "gravitational" mass of the galaxy, or the total of all stars, gas, and dark matter. The details of this calculation are described above in the Methodology section. What is important to note for this calculation, however, is that because our galaxies are unresolved, we are applying a generous definition of a rotation curve to resolve them. We are assuming that the rotational velocity of our galaxy can be taken as half of the velocity range that we had calculated and that the radius can be taken as half of the angular separation in the sky. By doing this, we have defined a rotation curve for our galaxies, and as such have "resolved" them to a certain extent. For iras 15250, specifically, we find that the actual dynamical mass is about $2.19 \times 10^9 M_{sun}$. With our calculated dynamical mass of $6.37 \times 10^9$, we find that our calculated dynamical mass has an error by a factor of about 3. It is encouraging that our calculation led us to a result that was not any orders of magnitude off the actual value. The error, then, can be attributed to our manual decision of which frequency bands showed emission. We mistakenly identified certain frequency bands as showing emission when they did not and should have instead been classified as background noise. This in turn would have led us to a narrower value for our velocity ranges and in turn, affected our dynamical mass calculations to be closer to the actual result.

The last step of our project was to determine the gas mass fraction of both galaxies. This was the ratio of molecular hydrogen to the total dynamical mass of the galaxy. Molecular hydrogen is the most abundant molecule in the universe, with it being about $10^4$ times as abundant as the CO molecule (taken from project description). There are several implications of the gas mass ratio. The gas mass ratio can tell us about the evolution of galaxies. Galaxies with high gas mass ratios typically have a lot of star formation left to complete. The abundance of molecular hydrogen gas makes it more likely for the gas clouds to condense, starting the chain of star formation. Conversely, galaxies with low gas mass ratios are less likely to have active star forming regions. With less molecular hydrogen available, there is far less chance of gas clouds collapsing under their own gravity and creating stars. Our galaxies each exhibit one of these traits. Iras 15250 has high gas mass ratio of 77%, meaning it is more likely to have active star forming regions and more stars will be born as the galaxy evolves. It also means as a whole, the galaxy is more dim (there is a direct correlation between luminosity and molecular hydrogen gas ratio). On the other side, iras 17208, with a gas mass ratio of 3.3% will have many stars already formed and be very luminous, more so than iras 15250. It is likely that iras 17208 is more along its galactic evolution than iras 15250 is.

5. DISCUSSION

Our biggest challenge with this project was understanding and manipulating our data. We observed images of two galaxies, iras 15250 and iras 17208, each with a length of about 1 arc-sec. Usually, image is not centered: fortunately, our image already was centered so we did not have to do any triple jump around our head. Initially, we obtained undeceive image that usually seen on TV, well known as microwave background. However, it is not something random. Our data consists of multiple polarization and spectral line channels, with each visibility having 112 channels. In order to separate background noise from the image of the galaxy we found the boundaries of the emission, by observing which channels had the greatest changes in average frequency.

In the case of iras 17208, there were a lot of different anomalies. We found that even if we selected 10 channels for observation, there was still an error among those channels. All of those channels are not necessary noise one ore few channels still could hold needed information. Due to nature of our project we did not include those channels. If we knew how to deal with the difmap we would include those individual channels. For example, lets say your boundary is [1,10) and on avg it is very small: however, there is a frequency spike in channel 5 in our case we would keep it as noise. Possible explanation would be, due to Earth rotation signal was received 100 percent precise and happened to be in our boundaries channel.

Another explanation for the noise could be the presence of a very bright star or star cluster, although this is unlikely because it implies that the star would be hundreds or thousand light years away from the galaxy. Due to these reasons, we used sections of channels rather than individual channels. We went through channels 11 and 99 for iras 17208 and found the x and y coordinates where the luminosity peaked for each set of 10 channels, in order to locate the brightest object in those channels. We used difmap the difmap command peak(x) and peak(y) to do this rather than generating plots of each set of channels and using our eyes, in order to eliminate the possibility of human error. Moreover, in case of iras 17208 the signal was very strong and its actual size could have be smaller than what we observed. When we generated UV-plot for iras 17208 using difmap, we found out we had linear lines in our plot rather than circular, which was expected because of our results during the tutorial. This can be attributed to multiple signals overlapping each other and causing artificial bright spots in our galaxy. Because of these plots we presumed to have a higher margin of error with iras 17208. Referring back to our difmap steps in 3, we observed where the location of the peak shifted, and found the shift in arcsec. We then used Ned Wright's Cosmology Calculator to find the angular distance using the redshift. We used a cosmology calculator here, since trigonometry will not work alone because we are dealing with red shift and we have to know the curvature of the universe for our values to be accurate, but we do not know it with certainty.

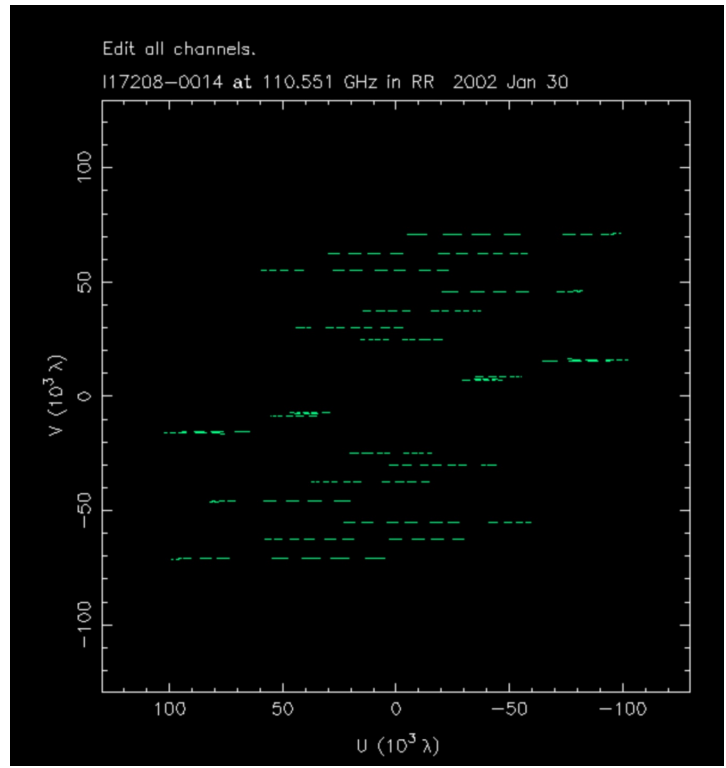

**Figure 6.** Visualization of uv array. In comparison iras 15250 looks more circular.

On uv-plot we can see that most of our data consist of straight lines. It is because the target is equatorial: as a result, uv-plot become linear It is good and bad news all together. it is good because we can say that we get very strong signal and did not have to deal with rotation of the Earth as much as we did for iras.15250: however, it cause different problems. Some arrays are overlap that cause significantly stronger/brighter regions on our target. In order to figure out if we have some sort of uncertainties we used vplot, projplot, radplot56, tplot.

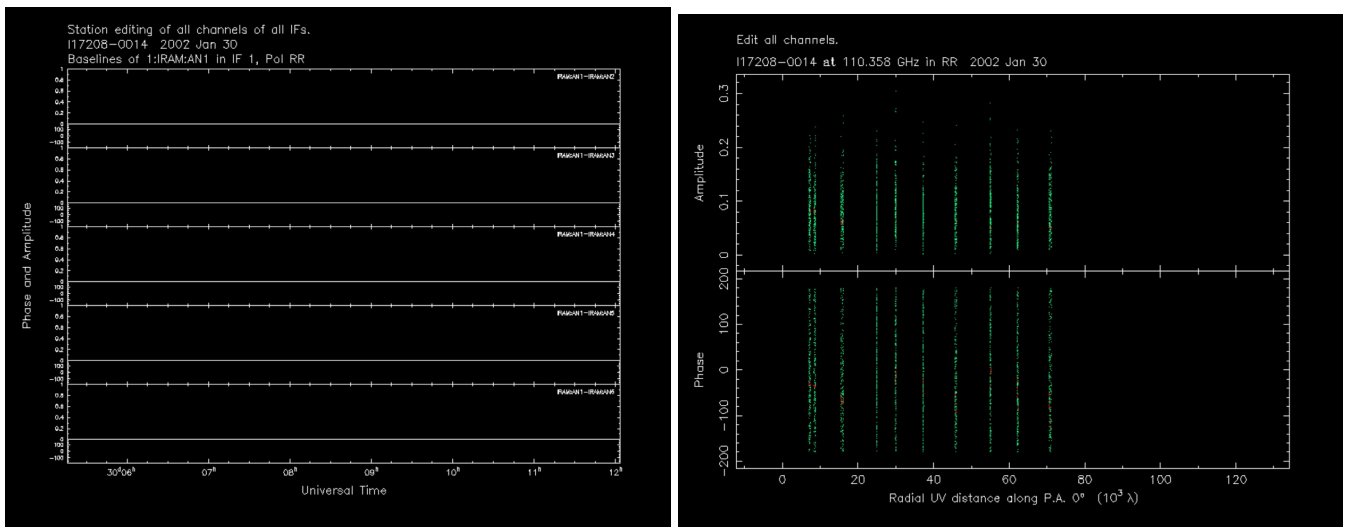

**Figure 7.** Visualization of v array and gives us better ideas of what is going on in Baseline 1 vs time

As we can see, vplot(on the left) graph is very linear. This plot combined all flaged data points from projplot and projplot 56 together. Typically we would press W and difmap would combine all flaged data points from projplot 56.

It is important to say that there are total 15 baselines and 6 stations. Moving forward, on the right side, we can see projplot. It does not allow us to identify radial UV-distance. The only thing that we may say that somewhere around 30 mas. data becomes predictable with same period between lines. It does not allow us to justify our results. Ideally we would want to see this data point distributed some where between 0 and 80 with some obviously more dens region. We can make any reasonable conclusions.

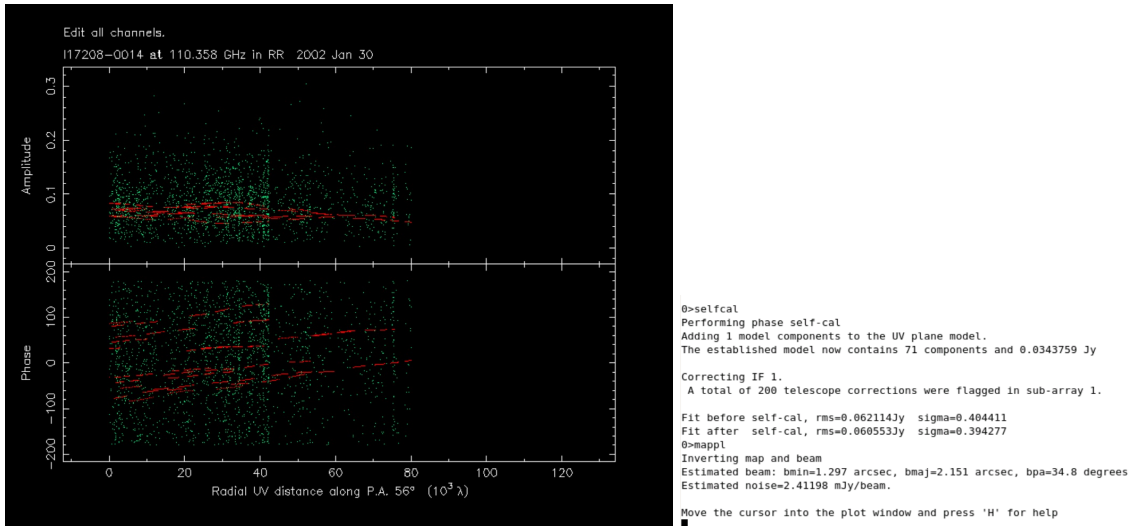

**Figure 8.** The result from projplot 56(on the left) for an extended double source with components along a position angle of 71. The beating of the two main component is not clearly seen and it is difficult to determine if wavelength directly related to their spatial separation of 10 or 62 or 71. On the right side is our final log that shows us that there is no errors in our cleaned data

Projplot 56 is more evenly distributed of projplot. It was final step in order to identify any potential errors in iras.17208 data set. In order to identify any errors in Basellines we are loking at the buttom of selfcal. As we can see there is nothing to be said about error such as "A total of 2 un-correctable visibility were flagged" difmap. It is reasonable to check for how long the following object was observed. To find we can use command spaceplot, it gives us time frame. In case of iras.17208 observation took place between 30/05:34:25 - 30/11:45:02 for total of 370.586 mins of observational time.

It would be scientifically interesting to also collect data from other wavelengths such as the X-ray or gamma ray range. Events such as supernovas or stellar objects such as pulsars can be detected in these ranges through their large amount of energy they give off at this range.

## 6. CONCLUSIONS

Our primary goal in this project was to quantify the total molecular masses of the galaxies, as well as their total gravitational masses. Through our research we gained an understanding of the roles that carbon (CO) and molecular hydrogen ($H_2$) play in terms of observing, measuring, and calculating these quantities. After making these calculations, we were able to draw the following conclusions about galaxy formation and evolution:

1. A high gas mass fraction indicates that the galaxy is young, and has consumed a small fraction of the gas so far. Particularly iras 15250 is a younger galaxy since it has used up less of its molecular gas mass whereas iras 17208 has used more of it indicating it is older.

2. Gas mass fraction and total gravitational mass can be used to calculate the amount of dark matter present in a galaxy.

3. Gas content and central surface brightness of the disk are directly correlated.

## 7. ACKNOWLEDGEMENTS

## REFERENCES

[1][1] Imanishi et al. *2004, ApJS 189, 270* *Alma Multiple-Transition Observations of High Density Molecular Tracers in Ultraluminous Infrared Galaxies*

[2][2] Lonsdale et al. *1 Mar 2006* *Ultraluminous Infrared Galaxies*

[3][3] Sorai et al. *Publi. Astron. Soc. Japan (2018)* *CO Multi-line Imaging of Nearby Galaxies (COMING)*

[4][4] SIMBAD Astronomical Database *SIMBAD Astronomical Database - CDS (Strasbourg)* *SIMBAD Astronomical Database*

[5][5] Rieke et al. *Draft version December 11, 2008* *Determining Star Formation Rates for Infrared Galaxies*